\documentclass[amsmath,amssymb,amsbsy,pre,twocolumn,tightenlines,superscriptaddress,floatfix]{revtex4-2}
\usepackage{amsfonts} % collection of miscellaneous TeX fonts
\usepackage{amsmath} 
\usepackage{amssymb}
\usepackage{blindtext}
\usepackage{lineno}
\usepackage{cancel}
\usepackage{bm}% bold math
\usepackage[usenames, dvipsnames]{color} % use colored tex
\usepackage{dcolumn}
\usepackage{epic} % enhanced picture environment
\usepackage{epsfig}
\usepackage{eucal} % changes the font selected by \mathcal
\usepackage{esdiff}
\usepackage{graphicx}
\usepackage[breaklinks,colorlinks = true,linkcolor = blue,urlcolor=blue,citecolor=red]{hyperref}

\begin{document}

%\title{Many-body Chaos in Fully Developed Turbulence}
\title{Intermittent fluctuations determine the nature of chaos in  turbulence}
\author{Aikya Banerjee}
\email{aikyabanerjee2016@gmail.com}
\affiliation{Department of Physical Sciences, Indian Institute of Science Education and Research Kolkata, Mohanpur 741246, India}
\affiliation{International Centre for Theoretical Sciences, Tata Institute of Fundamental Research, Bengaluru 560089, India}
\author{Ritwik Mukherjee}
\email{ritwik.mukherjee@icts.res.in}
\affiliation{International Centre for Theoretical Sciences, Tata Institute of Fundamental Research, Bengaluru 560089, India}
\author{Sugan Durai Murugan}
\email{vsdmfriend@gmail.com}
\affiliation{Department of Mechanical Engineering, Johns Hopkins University, Baltimore, Maryland 21218, USA}
\author{Subhro Bhattacharjee}
\email{subhro@icts.res.in}
\affiliation{International Centre for Theoretical Sciences, Tata Institute of Fundamental Research, Bengaluru 560089, India}
\author{Samriddhi Sankar Ray}
\email{samriddhisankarray@gmail.com}
\affiliation{International Centre for Theoretical Sciences, Tata Institute of Fundamental Research, Bengaluru 560089, India} 

	\begin{abstract}
We adapt recent ideas for many-body chaos in nonlinear, Hamiltonian fluids
	[Murugan \textit{et al.}, Phys. Rev. Lett. 127, 124501 (2021)] to
	revisit the question of the Reynolds number Re dependence of the Lyapunov exponent $\lambda\propto{\rm Re}^\alpha$
	in fully developed turbulence. The use of such decorrelators
	allow us to investigate 
	the interplay of the competing effects of viscous dissipation and nonlinearity. We obtain a precise value of $\alpha = 0.59 \pm 0.04$ and show that departure from the Kolmogorov 
		mean field result 
	$\lambda \propto \sqrt{{\rm Re}}$ is a consequence of the intermittent fluctuations in the velocity-gradient tensor. The robustness of our
	results are further confirmed in a local, dynamical systems model
	for turbulence.
\end{abstract}

\maketitle

Fully developed, incompressible turbulence is perhaps the most celebrated example of a chaotic
system. In contrast to other examples of classical, many-body
systems showing chaotic behaviour, turbulent flows have the distinction of
being central across natural world.
Unsurprisingly, therefore, 
physicists working in problems ranging from astrophysics, atmospheric sciences,
oceanography, and of course fluid dynamics have to factor in the underlying
chaotic nature of such flows~\cite{Pandit-Pramana-Review}. This ubiquity is not surprising: After all the
underlying Navier-Stokes equation, in all such systems, lead to solutions which
are turbulent in the limit of small viscosities commonly seen in most fluids~\cite{Frisch_1995}.

While mathematical and engineering tools remain indispensable, the ideas of
statistical physics provide the basis for much of our understanding of
turbulence, especially in its most general form where the flow is homogeneous
and isotropic. In particular, chaos along with the other fingerprint of fully
developed turbulence, namely intermittency, and their dependence \textit{inter
alia} the Reynolds number Re, forms a major challenge in a complete
understanding of turbulence. 

The \textit{degree} of chaos is usually quantified by the (largest) Lyapunov
exponent $\lambda$ of the flow. By using arguments tracing back to 
Kolmogorov's seminal work from 1941, Ruelle showed that 
$\lambda \propto \sqrt{{\rm Re}}$~\cite{Ruelle}. This result is a consequence of 
associating the (largest) Lyapunov exponent with the inverse of the smallest
time-scale of the flow  $\tau_\eta \propto 1/\sqrt{{\rm Re}}$, obtained most simply from 
phenomenological arguments, and thence the $\lambda$-scaling.
More generally, assuming \textit{a} H\"older exponent $h$ characterising the flow, 
it is easy to show $\lambda \propto Re^{\frac{1-h}{1+h}}$~\cite{Vulpiani}; indeed taking the \textit{monofractal},
Kolmogorov limit~\cite{K41} of a unique $h = 1/3$ recovers the Ruelle scaling. 

The result $\lambda \propto \sqrt{{\rm Re}}$ implicitly assumes a unique small length and
time-scale characteristic of a monofractal flow. However, the observational, experimental,
and numerical evidence against this is overwhelming. Indeed, a modern
rationalisation of turbulence rests on the multifractal approach~\cite{Benzi} developed by
Frisch and Parisi~\cite{Frisch-Parisi}. Adapting this model, which allows for a spread of H\"older
exponents $h$ in the flow, leads to the revised scaling obtained by Crisanti
\textit{et al.} $\lambda \propto {\rm Re}^\alpha$, with $\alpha \approx 0.459$~\cite{Vulpiani}.

This scaling exponent $\alpha \lesssim 0.5$ has been challenged in recent years
with data from different experiments and direct numerical simulations. For
example, Berera and Ho ~\cite{BereraPRL} have recently reported $\alpha =
0.53$. This lack of consensus between the theoretical estimate and measurements
--- as well as the lack of agreement between different
simulations~\cite{Yamada1987,Vulpiani,BereraPRL,Siddhartha,Boffetta_2017,Mohan_higher,Vassilicos_2023}
--- suggests that the origins of chaos in turbulence is far from settled.

But is there a way to address the microscopic origin of $\alpha$ from the
Navier-Stokes equation itself and connect it to the intermittent aspect of
fully developed turbulence?  And what role does the non-locality (in length
scale) of interactions --- a defining feature of the Navier-Stokes equation ---
play?

In this paper we show that this is indeed possible through recently
constructed ideas of \textit{decorrelators}~\cite{Lightcone} --- largely
limited to Hamiltonian~\cite{Subhro_PRB,Subhro_PRL,GTEuler} systems in the
context ergodicity, thermalisation and chaos --- and adapt them for
driven-dissipative, non-equilibrium systems such as turbulence. Consequently,
by using a suitable mix of theory and numerical simulations, we uncover the
competing effects of the viscous and nonlinear terms to show \textit{exactly}
the chaos and the scaling $\lambda \sim {\rm Re}^\alpha$ emerges along with
what determining the value of $\alpha = 0.59 \pm 0.04$ and why this
is so. We also confirm the robustness of this result by using a dynamical
systems approach to turbulence with the additional advantage of exploiting its
nearest and next-nearest neighbour interactions to investigate short-time
effects of locality which are absent in the Navier-Stokes equation.

We begin with the three-dimensional, incompressible, unit-density Navier-Stokes equation, driven by an external 
force to a non-equilibrium statistically steady state.
The force ${\bf f}$ varies to ensure a constant energy injection~\cite{Pope_forcing} $\mathcal{I}
= \langle {\bf f}\cdot {\bf u} \rangle \equiv \overline{\langle \varepsilon
\rangle}$, where the mean energy dissipation rate $\overline{\langle \varepsilon
\rangle}$ is the time $\overline{\cdot \cdot \cdot}$ and space $\langle \cdot
\cdot \cdot \rangle$ average of the spatio-temporally fluctuating dissipation
$\varepsilon ({\bf x},t)$ characteristic of such non-equilibrium steady states of fully developed 
turbulence with a statistically steady velocity field ${\bf u}_{\rm
steady}$. This follows from the familiar, instantaneous (spatially-averaged)
kinetic energy $E$ balance: $\frac{dE}{dt} = -\langle \varepsilon \rangle +
\mathcal{I}$ with a mean kinetic energy $E_0 = \overline {E}$.  

We now construct two, \textit{nearly} identical,  velocity fields ${\bf u_0}^{\rm A}
\equiv {\bf u}_{\rm steady}$ and ${\bf u_0}^{\rm B} = {\bf u_0}^{\rm A} +
\delta {\bf u_0}$, where the form of $\delta {\bf u_0}$ ensures that system B
remains incompressible.  We choose the amplitude of perturbation $\epsilon_0$
sufficiently small (see Appendix~\ref{App_DNS}) and localise it in space at ${\bf x}_0$; the precise choice 
of ${\bf x}_0$ is irrelevant and can be chosen to be the center of our domain.

We choose these fields ---  ${\bf u_0}^{\rm A}$ and  ${\bf u_0}^{\rm B}$ --- as
initial conditions $t = 0$ for the Navier-Stokes equation, evolve them independently in time, and
calculate the velocity difference $\delta {\bf u} ({\bf x},t) \equiv  {\bf
u}^{\rm B}({\bf x},t) - {\bf u}^{\rm A}({\bf x},t)$ for $t > 0$. From such
velocity differences, we are able to construct the decorrelator $\phi({\bf
x},t) \equiv \frac{1}{2} |\delta {\bf u} ({\bf x},t)|^2 $~\cite{GTEuler,mukherjee2023intermittency}.
It is easy to show, starting from the Navier-Stokes equation written separately for systems A and
B, that the decorrelator follows an evolution (written in component form and 
sum over repeated indices assumed)
\begin{equation}
	\frac{\partial \phi(\mathbf{x},t)}{\partial t} = \partial_i W_i - \delta u_i S_{ij} \delta u_j + 
	\nu \delta u_i \nabla^2\delta u_i
	\label{eq:phi}
\end{equation}
with the strain-rate tensor 
$S_{ij} = \frac{1}{2}(\partial_i u^A_j + \partial_j \delta u^A_i)$,  
$W_i = u^B_i \phi + \delta u_i \partial_{jk}\int \mathrm{d}^3 \mathbf{x'} 
G(\mathbf{x},\mathbf{x'})[u^A_j\delta u_k + u^A_k \delta u_j + \delta u_j \delta u_k]'$, 
where $G(\mathbf{x}, \mathbf{x'})$ is the Green's function obtained from the Poisson equation for pressure, and $\nu$ the coefficient of
kinematic viscosity (see Appendix~\ref{Phi_derivation}).

A further simplification exploits the statistical isotropy and
homogeneity of fully developed turbulence to construct, over the volume
$\mathcal{V}$, the spatially integrated decorrelator $\Phi (t) \equiv
\frac{1}{\mathcal{V}}\int \mathrm{d}{\bf x} \phi ({\bf x},t) = \langle \phi
\rangle$. This leads to a simple cancellation of the divergence term in
Eq.~\eqref{eq:phi} and a resulting evolution equation for the integrated decorrelator:

\begin{equation}
	\dot{\Phi} = \frac{d\Phi}{dt} = \beta_S + \beta_\eta,
	 \label{eq:Phi}
\end{equation}
where $\beta_S = -\langle \delta {\bf u}\cdot \mathbf{S} \cdot \delta {\bf u}\rangle$ and $\beta_\eta = \nu \langle \delta {\bf u} \cdot \nabla^2 \delta {\bf u}\rangle$ as 
the contributions from the strain and dissipative terms, respectively.
There is one additional remark to make.  In the derivations of Eqs.~\eqref{eq:phi}-\eqref{eq:Phi} we have neglected 
the forcing terms on systems A and B. This is perfectly valid at short times but in the long time $t \to \infty$ limit 
it plays a critical role as we shall see below.

At short times we conjecture an exponential growth of the decorrelator: 
$\dot{\Phi}/\Phi \equiv \lambda$.
We confirm this numerically by integrating Eq.~\eqref{eq:Phi} with an initial perturbation 
field $\delta {\bf u}$ and measurements of the strain-rate tensor $\mathbf{S}$ drawn from direct numerical 
simulations (DNSs) of the Navier-Stokes equation (Appendix~\ref{App_DNS}) 

\begin{figure}
\includegraphics[width = 1.00\linewidth]{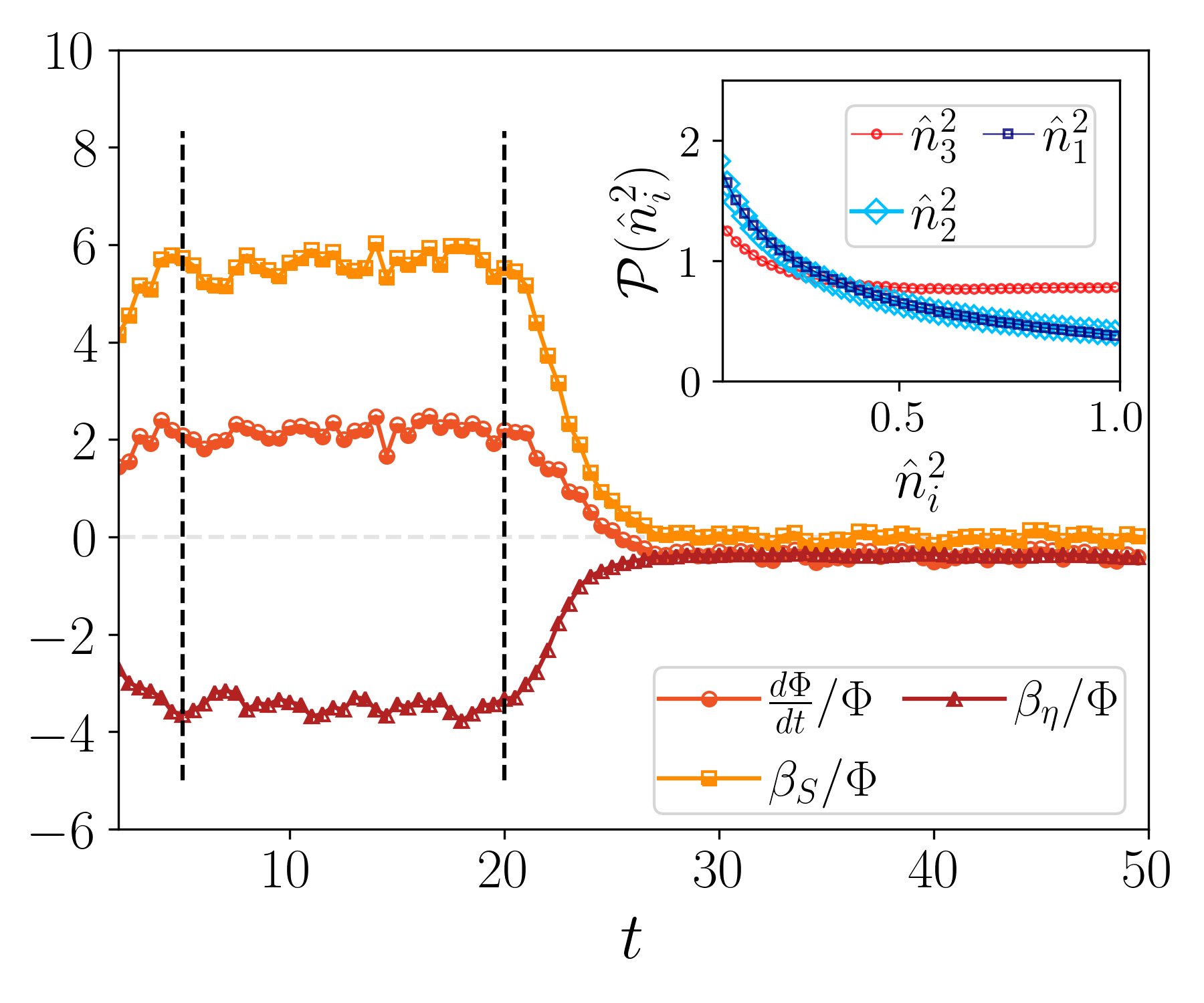}
	\caption{Representative plots of the normalised spatially averaged decorrelator $\dot{\Phi}/\Phi$,  
	 $\beta_S/\Phi$ 
	and $\beta_\eta/\Phi$ vs time for $Re = 178.5$. The pair of vertical dashed lines 
	indicate the short-time exponential-growth phase.
	(Inset) Probability density 
	function of the ${\hat{n}_i}^2$ direction cosines of ${\bf u}$ (at short times) showing
	a preferential growth of the difference field along the 
	compressional $i = 3$ direction.}
	\label{fig:Phi-theory}
\end{figure}

In Fig.~\ref{fig:Phi-theory} we show a representative plot of $\dot{\Phi}/\Phi$
vs time for ${\rm Re} = 178.5$ displaying a clear plateau $\lambda$
demarcated by a pair of vertical dashed lines. While this is not very surprising, 
given that turbulence is chaotic, it is far from obvious how to trace its origins in
the structure of Eq.~\eqref{eq:Phi} and in particular how the competing effects of
the strain and viscous terms conspire to give such exponential-%\textit{scrambling}
growth phases.
In Fig.~\ref{fig:Phi-theory} we plot $\beta_S$ and $\beta_\eta$ (compensated by $\Phi$) and find strong
evidence of their short-time exponential growth --- the plateau between the
vertical dashed lines --- corresponding to Lyapunov exponents $\lambda_S$ and
$\lambda_\eta$, respectively. Furthermore, the dissipative effect of the
viscous term $\lambda_\eta \sim \mathcal{O}(1/\tau_\eta) < 0$ is compensated by the
effect of the strain term $\lambda_S > |\lambda_\eta|$, leading to an overall positive
Lyapunov exponent $\lambda = \lambda_S + \lambda_\eta$. 

How do we understand the origins of $\lambda_S$ which leads to a compensation
of the negative $\lambda_\eta$ and hence a positive $\lambda$? In particular,
how does the strain field $S$ ensure the exponential amplification of $\delta
{\bf u}$ at short times and thence $\Phi \sim e^{\lambda t}$? The answer lies
in decomposing the strain term in the eigenbasis of the strain tensor $
\delta {\bf u}\cdot \mathbf{S} \cdot \delta {\bf u} = \sum_{i = 1}^{i=3}
\hat{n}_i^2\gamma_i |\delta {\bf u}|^2$, with ${\hat{n}_i}$
direction cosines of ${\bf u}$ (along the eigendirections) and ${\gamma_i}$ the
eigenvalues. The relative distributions for the extensional ($i = 1$;
$\gamma_1 > 0$), intermediate  ($i = 2$; $\gamma_2 \approx 0$) and
compressional directions ($i = 3$; $\gamma_3 < 0$), seen in the inset of
Fig.~\ref{fig:Phi-theory}, underlines a bias for the compressional direction. 
Thus, clearly the dominant contribution in this expansion \textit{must}
come from the statistics of the  compressional eigenvalue.  

At such short times, a careful analysis of $\beta_S$ shows (see Appendix~\ref{Phi_derivation}) that 
$\beta_S \sim \langle \gamma_3 \rangle + \gamma_3^{\rm std}$. And hence, the Reynolds number 
dependence of $\lambda_S$ must be dominated by the competing effects of the mean $\langle \gamma_3 \rangle$ 
eigenvalue and their intermittency-induced fluctuations $\gamma_3^{\rm std}$.

In the long time limit, $\beta_S \to 0$, all eigenvalues are sampled
equally (see Appendix~\ref{Phi_derivation}, Fig.~\ref{3fig}(c)) leading to  $\langle \delta {\bf u}\cdot \mathbf{S}
\cdot \delta {\bf u}\rangle \propto \sum_{i = 1}^{i=3} \gamma_i = 0$ because of
incompressibility. We see from Fig.~\ref{fig:Phi-theory} that this is indeed
the case as the compensated $\beta_S$ term rapidly falls to 0 for $t \gtrsim 25$.

Does the viscous term $\beta_\eta$ also go to zero similarly at late times
leading to $\dot{\Phi} = 0$? 
At such late times it is easy to show that (see Appendix~\ref{Phi_derivation})
$\beta_\eta \to -2\overline{\langle \varepsilon\rangle}$. Indeed, in
Fig.~\ref{fig:Phi-theory} we do see a clear saturation of the normalised
viscous term to a value consistent with $-2\overline{\langle \varepsilon\rangle}
\approx 1.6$ (corresponding to the simulations from which the rate of strain
matrix is drawn).

A trivial consequence of this is that within the framework of
Eq.~\eqref{eq:Phi} $\dot{\Phi} = -2\overline{\langle \varepsilon\rangle}$ (clearly
seen in Fig.~\ref{fig:Phi-theory}) and suggesting that the decorrelator never
saturates. However this is a contradiction: By definition, the decorrelator is
defined as the spatial average of the velocity difference of systems A and B,
and hence at long times $\Phi \to 2E_0$ since the cross-correlator $\langle
{\bf u}_{\rm A} \cdot  {\bf u}_{\rm B}\rangle$ vanishes as the fields
decorrelate as $t \to \infty$.

This contradiction is resolved by recalling that the derivation of
Eq.~\eqref{eq:Phi} neglects the effective energy injection. While at short
times, this is zero, at long times this contribution is no longer irrelevant.
In fact we can show (see Appendix~\ref{Phi_derivation}) that $\langle \delta {\bf f}\cdot \delta
{\bf u} \rangle = 2\overline{\langle \varepsilon\rangle}$ as $t\to \infty$ and thus
compensates the viscous contribution $-2\overline{\langle \varepsilon\rangle}$ (see Appendix~\ref{Phi_derivation}, Fig.~\ref{3fig}(a))
leading to $\dot{\Phi} = 0$.

\begin{figure}
\includegraphics[width = 1.00\linewidth]{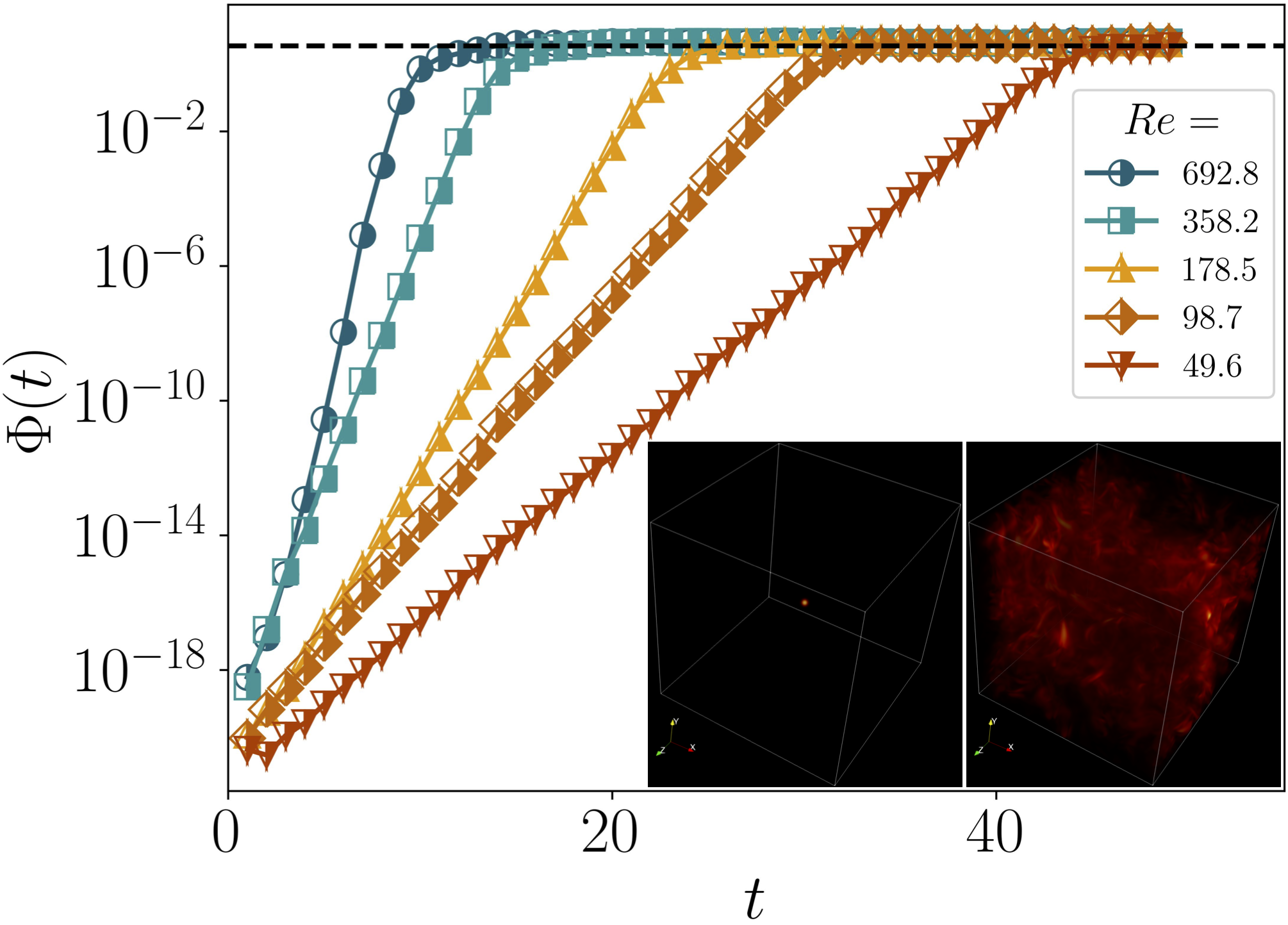}
	\caption{Semilog plots of $\Phi$ vs $t$ from DNSs for different Re showing an exponential regime followed by 
	saturation at a value $2E_0$ (dashed horizontal line).
	The insets show two snapshots at (left) $t = 0$  and (right) $t = 12.5$ of the square of the difference field
	$|\delta {\bf u}|^2$ for Re = 178.5. (See Ref.~\cite{movie} for the full evolution of $|\delta {\bf u}|^2$.)}
	\label{fig:DNS}
\end{figure}

How consistent are these ideas in actual measurements of the decorrelator in DNSs of the 
Navier-Stokes equation? We check for this by solving the incompressible Navier-Stokes equation 
on a 2$\pi$ triply-periodic domain by using a pseudospectral method with a large-scale constant energy $\mathcal{I} = \overline{\langle \varepsilon\rangle}$~\cite{Pope_forcing}
injection scheme to drive the flow and maintain it in non equilibrium statistically steady state with a 
mean energy $E_0$; see Appendix~\ref{App_DNS} 
for a detailed summary of the numerical scheme and the parameters used. 
By using the same protocol described earlier, we calculate the decorrelator $\Phi$ for several 
different values of the %Taylor-scale
Reynolds number ${\rm Re}$. In
Fig.~\ref{fig:DNS} we show representative snapshots 
at $t = 0$ (left inset) and $t = 12.5$ (right inset), corresponding to the initial and 
exponential-growth phases, respectively. (Ref.~\cite{movie} shows the full evolution of $|\delta {\bf u}|^2$.)

These difference fields allow us to construct the spatially averaged
decorrelator $\Phi$ which we plot, on a semilog scale,  in the main panel of
Fig.~\ref{fig:DNS} as a function of time. We see a convincing exponential regime 
from which we can extract the Lyapunov exponent $\lambda^{\rm DNS}$, followed by a saturation at a value approximately 
equal to $2E_0$, indicated by the dashed horizontal line, as our theory suggests. (We have confirmed from our data that the 
onset of the saturation of the decorrelators (see Figs.~\ref{fig:Phi-theory} and~\ref{fig:DNS}) is of the order of the 
inverse of the Lyapunov exponent.) 

It is important to observe that even at very short times, before the exponential growth phase, 
our results from the Navier-Stokes equation summarised in Fig.~\ref{fig:DNS} do not suggest an initial 
power-law growth which has been seen in many local models of chaotic systems\cite{Luca_shell, Ditlevsen-book}. Could this --- as well as the
somewhat self-similar evolution of the difference field (see Fig.~\ref{fig:DNS}, insets) --- be a consequence 
of the essential non-local nature of the equation? 

\begin{figure}[b!]
\includegraphics[width = 1.00\linewidth]{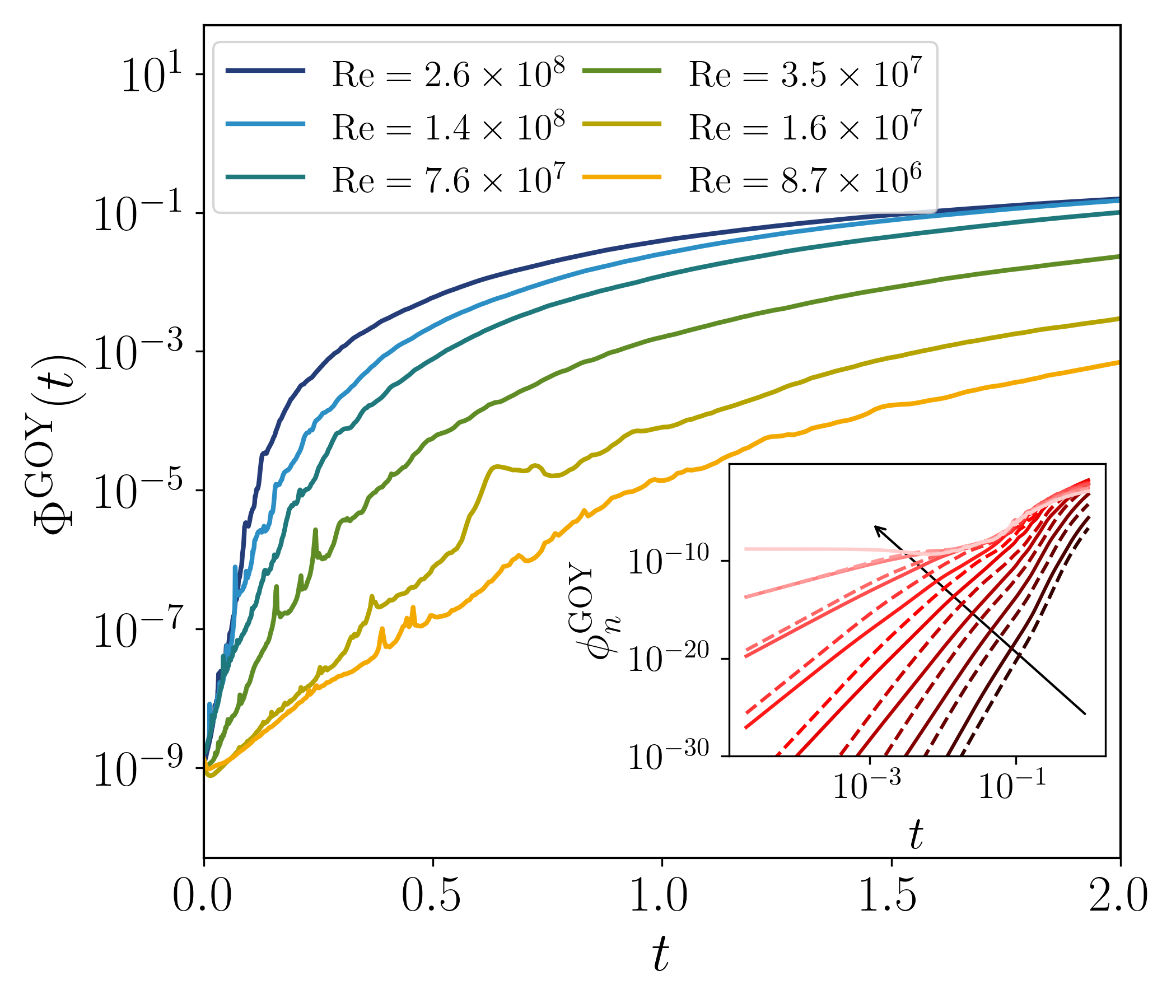}
\caption{Semilog plot of the shell model decorrelator $\Phi^{\rm GOY}$ vs $t$ for different Re showing an exponential growth.
	(Inset) Loglog plot of the shell-specific decorrelator $\phi_n$ for different shells (the arrow indicates the 
	direction $n = 1$ to $n = 16$ shells) at \textit{very} early times shows a power-law growth.}
	\label{fig:shell-decorr}
\end{figure}

To test this as well as the robustness of our conclusions of an exponentially growing decorrelator, 
we use a simpler, phenomenological (cascade) model for turbulence, inspired from dynamical systems, 
namely the Gledzer-Ohkitani-Yamada (GOY) shell model~\cite{GOY-Gledzer, GOY-Okhitani, Ditlevsen-book}.
In such shell models (Appendix~\ref{App_DNS}), the velocities $u_n$, associated with a scalar, exponentially
growing, wave-number $k_n = k_0 2^n$, are treated as complex, dynamical
variables, with the shells numbers being integers $1 \le n \le N$ (see Appendix~\ref{App_DNS} for further details). 

\begin{figure}
\includegraphics[width = 1.00\linewidth]{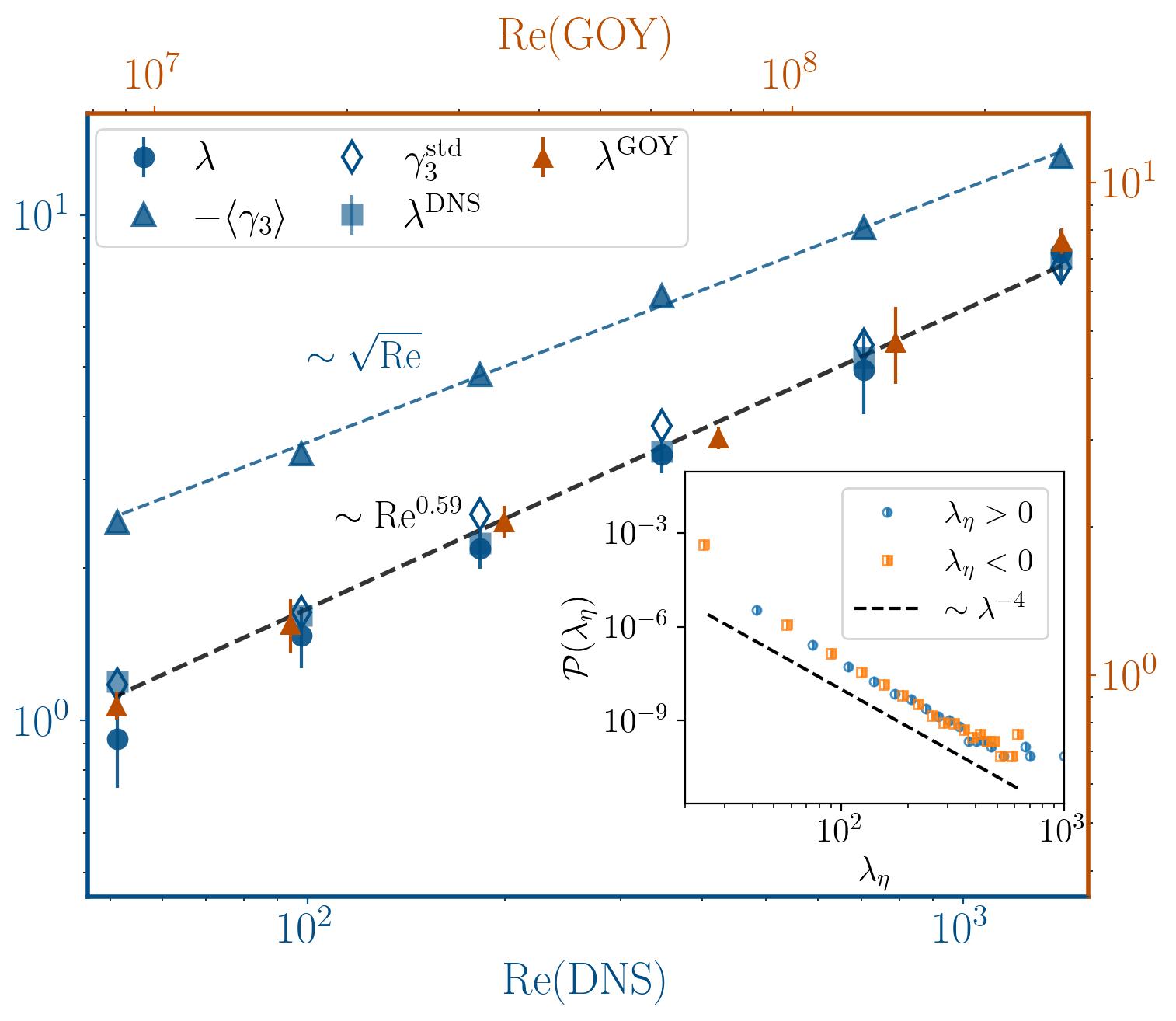}
	\caption{A loglog plot of $\lambda$, $\lambda^{\rm DNS}$, $\gamma_3^{\rm std}$
	and $-\langle \gamma_3 \rangle$, obtained from various DNSs versus the Reynolds number Re (DNS) (bottom X axis), 
	as well as the Lyapunov exponent obtain from the shell model $\lambda^{\rm GOY}$ (rescaled by a constant factor 
	to overlap with the DNS data since the Reynolds number ranges are different) with the 
	corresponding shell model Reynolds numbers Re(GOY) (top X axis). 
	(Inset) A loglog plot of the probability density function $\mathcal{P}$ for 
	$\lambda_\eta$ showing a power-law $|\lambda_\eta|^{-4}$ tail.}
	\label{fig:lambda}
\end{figure}

In the statistically steady state (Appendix~\ref{App_DNS}), let the velocity \textit{field} be denoted as $u_n^{\rm A}$. We 
now construct a second field $u_n^{\rm B} = u_n^{\rm A} \forall n \neq n_p$ and 
$u_{n_p}^{\rm B} = (1 + \epsilon_0) u_{n_p}^{\rm A}$. Our choice of $n_p = 16$ ensures 
that the tiny perturbation (with $\epsilon_0 = 10^{-3}$) are always added at wavenumbers just before
the onset of the dissipation scales for all the Reynolds number being considered. (We have checked that our results
are robust to the specific choice of $n_p$ or variations in $\epsilon_0$.)

We now use the nearly-identical fields $u_n^{\rm A}$ and $u_n^{\rm B}$ as
initial conditions and evolve the GOY shell model independently for systems A and B
and hence construct the shell model decorrelator $\phi_n (t) \equiv \langle |u_n^{\rm A} - u_n^{\rm B}|^2 / 2\rangle$
The $\langle \cdot \rangle$ denote an ensemble averaging over independent realisations of the initial field
$u_n^{\rm A}$. While it is straightforward to write the evolution equation for $\phi_n$ (Appendix~\ref{App_shell}), we 
find this is neither revealing nor analytically 
tractable. Instead, we choose to work with the numerically constructed $\phi_n$ and assess its behaviour from 
our simulations.  

Given the local nature of interactions in the shell model\cite{Luca_shell}, it is reasonable to
expect that decorrelators such as $\phi_n$ may well have a self-similar short
time growth followed by an eventual exponential rise with a saturation as $t
\to \infty$. In the inset of Fig.~\ref{fig:shell-decorr} we show a loglog
plot of the temporal evolution of $\phi_n$ for a small initial time window $0
\le t \lesssim 0.5$ with a clear power-law range for $n < n_p$. This power-law eventually ($t \gtrsim 0.1$) gives way
to the %\textit{scrambling} phase~\cite{Shenker2014} marked by an
exponential-growth phase of
decorrelators. The
Lyapunov exponent characterising such shell-by-shell decorrelators $\phi_n$ are
of course $n$-dependent. Hence we find it useful to construct the $n$-independent Lyapunov exponent $\lambda^{\rm GOY}$ 
via the integrated decorrelator $\Phi^{\rm GOY} = \sum_1^N \phi_n$.
In the main panel of Fig.~\ref{fig:shell-decorr} we show a semilog plot of the
temporal evolution of $\Phi^{\rm GOY}$ for different values of Re
showing, yet again, a clear exponential growth, with a shell model Lyapunov exponent 
$\lambda^{\rm GOY}$, followed by a saturation to
values roughly corresponding to $2E_0$, with $E_0$ the statistically
steady state energy in our shell model.

We can now summarise all of these ideas in answering the
question of how consistent are the various (largest) Lyapunov exponent ---
$\lambda$, $\lambda^{\rm DNS}$, and $\lambda^{\rm GOY}$ --- with each other given the 
different ways and indeed the different models from which they are derived?
In Fig.~\ref{fig:lambda} we show a composite plot of $\lambda$, $\lambda^{\rm
DNS}$ and $\lambda^{\rm GOY}$ versus the 
Reynolds number Re. For measurements from the DNSs and the GOY shell model
simulations, we estimate the Lyapunov exponents from a linear fit in the
exponential-growth %scrambling
phase of the decorrelators (as shown in Figs.~\ref{fig:DNS} and
\ref{fig:shell-decorr}); the errors on such fits yield the errorbars on
$\lambda^{\rm DNS}$ and $\lambda^{\rm GOY}$. We find our errorbars smaller than
the symbols sizes in Fig.~\ref{fig:lambda}. The errorbars for the exponent
$\gamma_3$ are extracted likewise and are of the same order as those on
$\lambda^{\rm DNS}$ and $\lambda^{\rm GOY}$.  The exponent $\lambda$ is
obtained differently. We use plots of $\dot{\Phi}/\Phi$, such as the one shown
in Fig.~\ref{fig:Phi-theory}, to extract $\lambda$ and its errorbar as the mean
and standard deviation of the plateau, respectively.

We find remarkable consistency (within error bars) between the different
measurements of the Lyapunov exponents and indeed across the different models,
as well as the range of Reynolds number between the DNSs (bottom axis) and
shell models (top axis), that we use. All the Lypanov exponents seem to follow
a universal scaling: $\lambda^{\rm DNS}$ = $\lambda^{\rm GOY}$ = $\lambda$
$\sim {\rm Re}^{\alpha}$, with $\alpha = 0.59 \pm 0.04$. Remarkably, combining the 
shell model and DNS data we show in Fig.~\ref{fig:lambda} that our reported scaling 
holds for nearly 7 decades in Reynolds number.

The theory developed earlier suggests a connection between the Lyapunov exponent and the statistics of 
$\gamma_3$ with $\lambda_S \sim \langle \gamma_3 \rangle + \gamma_3^{\rm std}$ (Appendix~\ref{Phi_derivation}). 
The departure of the measured Lyapunov exponents (Fig.~\ref{fig:lambda}) from a Kolmogorov-like mean field result $\lambda \sim \langle \gamma_3 \rangle  \sim \sqrt{{\rm Re}}$, is a clear indication that the 
deviation must stem from intermittent fluctuations and the dominant scaling in the statistics of $\gamma_3$ is 
due to $\gamma_3^{\rm std} \sim {\rm Re}^\alpha$. In Fig.~\ref{fig:lambda} we show a plot of $\gamma_3^{\rm std}$ from our DNSs and find a near perfect agreement with the Lyapunov exponent confirming, nearly 7 decades in Reynolds number, the 
robustness of this theory: $\lambda = \lambda^{\rm DNS} = \gamma_3^{\rm std} \sim {\rm Re}^{0.59}$. Furthermore, 
we confirm that the mean eigenvalue -$\langle \gamma_3 \rangle \sim \sqrt{{\rm Re}}$ (Fig.~\ref{fig:lambda}) consistent with the simpler scaling analysis, not accounting 
for fluctuations, and at odds with the measured Lyapunov exponent.

In this work we have shown how recent ideas of decorrelators~\cite{Subhro_PRL, Subhro_PRB,Lightcone,Shenker2014, GTEuler} 
provide a microscopic way to understand the origins of chaos in fully developed turbulence.
Remarkably, these ideas are just as robust for simpler, cascade models of turbulence where
locality~\cite{Luca_shell} ensures an 
initial scale-invariant growth of the decorrelator.
We find signatures of these fluctuations in one further aspect of the strain
and viscous terms in Eq.~\eqref{eq:Phi} which connects with the idea of
intermittency. The theory developed by us involve the
\textit{mean} exponents: $\lambda_S$ and $\lambda_\eta$ (and hence $\lambda$) 
are obtained from the spatial integrals over the corresponding terms.
However, it is possible to measure the \textit{local} exponents
$\lambda_S ({\bf x}) \equiv  \frac{\delta {\bf u}\cdot \mathbf{S} \cdot \delta {\bf
u}}{|\delta {\bf u}|^2/2}$ and $\lambda_\eta ({\bf x}) \equiv \frac{\nu \delta {\bf u}
\cdot \nabla^2 \delta {\bf u}}{|\delta {\bf u}|^2/2}$ in the exponential-growth 
phase
and thence their probability distribution functions $\mathcal{P}$.  These 
distribution are found to have
exponential (for $\lambda_S ({\bf x})$) or power-law (for $\lambda_\eta ({\bf x})$; see inset of 
Fig.~\ref{fig:lambda}) tails strongly suggestive of intermittency and its connections with the question of 
chaos in turbulence. 
We do not explore these ideas, and in particular
the origins of the power-law tails, any further in this work as well as its implications for Lagrangian chaos~\cite{Jeremie_Boffetta_Lagrangian_tracers,Ray2018}, but leave it for a more detailed 
study of the statistics of $\beta_S$ and $\beta_\eta$ in the future.

\begin{acknowledgements}
        We acknowledge Anupam Kundu, Siddhartha Mukherjee, Sthitadhi Roy, Sibaram Ruidas and Akash Sarkar for discussions on related issues.
	A.B. acknowledges the Long Term Visiting Students' Program (LTSVP) of the 
	ICTS-TIFR which enabled this collaboration. 
	SB acknowledges funding from the Swarna Jayanti fellowship of SERB-DST (India) Grant
	No. SB/SJF/2021-22/12 and DST, Government of India (Nano mission), 
	under Project No. DST/NM/TUE/QM-10/2019 (C)/7.
	S.S.R. acknowledges SERB-DST (India) projects 
	STR/2021/000023 and CRG/2021/002766 for financial support. 
	This research was supported in part by the International Centre for Theoretical Sciences (ICTS) for the programs - 
	Field Theory and Turbulence ( (code:ICTS/ftt2023/12), 
	Indo-French workshop on Classical and quantum dynamics in out of equilibrium systems  (code: ICTS/ifwcqm2024/12) and 
        10th Indian Statistical Physics Community Meeting (code: ICTS/10thISPCM2025/04). 
	The simulations were performed on the ICTS clusters Mario, Tetris, and Contra.
	The authors 
	acknowledge the support of the DAE, Government of India,
	under projects nos. 12-R\&D-TFR-5.10-1100 and RTI4001.
\end{acknowledgements}

%%%%%%%%%%%%%%%%%%%%%%%%%%%%%%%%%%%%%%%%%%%%%%%%%%%%%%%%%%%%%%%%%%%%%%%%%%%%%%
\bibliographystyle{apsrev4-2}  % This ensures the REVTeX bibliography style
\bibliography{decorrelator140525} 
\appendix
\onecolumngrid

\section{Direct Numerical Simulations}
\label{App_DNS} 

\renewcommand{\theequation}{A-\arabic{equation}}
\setcounter{equation}{0}  % reset counter 

\subsection{The Navier-Stokes Equation}

We perform direct numerical simulations (DNSs), by using a pseudospectral method, of the three-dimensional (3D), 
incompressible, unit density Navier Stokes equation, with the pressure field $P$
\begin{equation}
	\frac{\partial {\bf u}}{\partial t} + {\bf u}\cdot\nabla {\bf u} = \nu \nabla^2 {\bf u} - \nabla P + {\bf f}.
\label{eq:NS}
\end{equation}
on a periodic box of size $L = 2\pi$ with $N^3$ collocation points. We choose $N = 256$ and $N = 512$ (to check for numerical 
convergence) and vary the coefficient viscosity $10^{-3}\le \nu \le 32\times10^{-3}$ to obtain Reynolds numbers $ 50 \le {\rm Re} \le 1400 $. 
We use a second-order 
Adams-Bashforth for time-marching with a time step $\delta t = 5  \times 10^{-4} $ (for $N = 256$) and 
$\delta t = 4\times 10^{-4} \ $ (for $N = 512$).

We initialise the flow with a random initial condition such that the  
initial energy spectrum spectrum $E(k) = A_0 k^2 \exp(-k^2 / 2k_0^2)$ with $A_0
= k_0 = 1$. The system is driven to a non-equilibrium steady state (NESS)
through a large-scale forcing with a constant energy injection at large scales
corresponding to wavenumber $1 \le k_{\rm force}\le 2$. 

This statistically steady velocity field ${\bf u}$ is taken as the initial field for system A: ${\bf u_0}^{\rm A}
\equiv {\bf u}$ and further define the initial condition for system B via ${\bf u_0}^{\rm B} = {\bf u_0}^{\rm A} +
\delta {\bf u_0}$. We use a Gaussian perturbation
$\delta {\bf u}_0 ({\bf x}) = \epsilon_0 \exp{(-(x - x_0)^2 / 2\sigma_0^2)}{\bf \hat{x}}$ with $\epsilon_0 = 10^{-5}$, $\sigma_0=4{\rm d}x $ and $x_0 = ({\pi, \pi, \pi})$ at the center of the domain and which also satisfies the incompressibility constraint.

The nearly identical fields  ${\bf u_0}^{\rm A}$ and  ${\bf u_0}^{\rm B}$ are then used as initial conditions for the Navier-Stokes 
equation and evolved as before. 
The decorrelator is calculated from the evolution of such twin simulations as 
$\phi({\bf x},t) = |{\bf u}^{\rm B} - {\bf u}^{\rm A}|^2/2$ and thence $\Phi(t) = \int d{\bf x} \phi({\bf x},t) / \mathcal{V}$. 
The decorrelators are then used to estimate the Lyapunov exponents (obtained by fitting the initial exponential growth phase) 
and its dependence on the Reynolds number.

\subsection{The Gledzer-Ohkitani-Yamada (GOY) Model}
The Gledzer-Ohkitani-Yamada or the GOY shell model
\begin{eqnarray}
	\frac{d}{dt} u_n &=& \iota k_n [u_{n+2}u_{n+1} - \frac{1}{4}u_{n+1}u_{n-1}
	- \frac{1}{8} u_{n-1}u_{n-2}]^* \nonumber \\ &-& \nu k_n^2 u_n + f
	\label{eq:GOY}	
\end{eqnarray}
is one of several cascade models which mimic Navier-Stokes turbulence and are designed to achieve 
very high Reynolds number. 
In such models, the velocities $u_n$ (with boundary conditions $u_{-1} =
u_{0} = u_{N+1} = u_{N+2} = 0$)  associated with a scalar, exponentially
growing, wave-number $k_n = k_0 2^n$, are complex, dynamical
variables. The shell numbers range from $1 \le n \le N$; in our simulations we
choose $N = 22$ and the coefficient of viscosity varies as $2\times 10^{-7} \le
\nu \le 64 \times 10^{-7}$ yielding Reynolds numbers in the range $ 8.7 \times 10^6
 \le {\rm Re}
\le 2.6 \times 10^8$; the shell model, %Taylor-scale
Reynolds number is simply defined as ${\rm Re} = |u_{\rm rms}|/k_0\nu$.
The equation is integrated numerically by using a second-order Runge-Kutta scheme with a time step $\delta t = 2 \times 10^{-5}$
The constant amplitude forcing $f = 0.1$, applied on the $n = 2$  shell and drives the system to a statistically steady 
state~\cite{Ray2008}.

As with the strategy for the Navier-Stokes equation, we first obtain a non-equilibrium statistically steady 
state with  an energy spectrum $E(k_n) \equiv |u_n|^2/k_n \sim k_n^{-5/3}$ over an inertial range 
of shells whose extent is determined by Re. We use this steady state velocity field to define the initial 
field for system A --- $u_{n,0}^{\rm A}$ and construct the initial condition for system B via
$u^B_{n,0} = (1 + \epsilon_0)
\delta_{n,n_p} u_{n,0}^A$. In all our simulations we have used $\epsilon_0 =
0.001$ and $n_p = 16$.  Numerically it has been seen that the shell wise
decorrelators first show an power-law rise later followed by the
exponential-{growth}%scrambling
regime. The initial power law largely depends on the
distance from the perturbed shell. Therefore, we add the perturbation in a fixed
shell $n_p = 16$ for all values of ${\rm Re}$.

% =============================================================== %
% Appen - B
% =============================================================== %

\section{Derivation of the decorrelator from the Navier-Stokes equation and its limits} \label{Phi_derivation}
\renewcommand{\theequation}{B-\arabic{equation}}
\setcounter{equation}{0}  % reset counter 
\subsection{Evolution equation for $\phi({\bf r},t)$ and $\Phi$}
In this section we provide the derivation of the equation of motion of the decorrelator for the Navier-Stokes equation and 
in particular estimate the long and short time limits of $\beta_S$ and $\beta_\eta$.

We begin by recalling that the pressure term in the Navier-Stokes equation can be rewritten as a Poisson equation  
\begin{equation}
\nabla^2 P = - {\partial_j u_i}{\partial_i u_j}
\end{equation}
by exploiting the incompressibility constraint. This Poisson equation underlines the non-local nature of turbulence.

It is useful, in what follows, to have a Green's function representation
\begin{equation} 
	P({\bf x}) = -\int {\rm d}{\bf x} G({\bf x},{\bf x}^\prime)
	\left [{\partial_j u_i}{\partial_i u_j}\right ]^\prime 
\end{equation}
	where the $\prime$ denotes the quantity evaluated at ${\bf x}^\prime$.

This allows us to construct the 
evolution equation of the solenoidal 
velocity perturbation $\delta {\bf u} = {\bf u}^{\rm B} - {\bf u}^{\rm A}$ in component form 
\begin{equation}
\frac{\partial \delta u_i}{\partial t} = - \delta u_j \partial_j u_i^{\rm A} - u_j^{\rm B} \partial_j \delta u_i
-  \partial_i \left(\int {\rm d}{\bf x}
G({\bf x},{\bf x'}) [\partial^2_{jk} (u_j^{\rm A}\delta u_k +\delta u_j u_k^{\rm A} + \delta u_j\delta u_k)]'\right) + \nu \nabla^2 \delta u_i + \delta u_i \delta f_{i} 
\label{delu}
\end{equation}
and hence, via a dot product of  $\delta {\bf u}$ with Eq.~\eqref{delu}, the equation for 
the decorrelator $\phi ({\bf r},t)$:
\begin{equation}
\frac{\partial \phi}{\partial t} = - \delta u_i S_{ij} \delta u_j 
-  \partial_i W_i + \nu \delta u _i\nabla^2 \delta u_i + \delta u_i \delta f_{i}.
\label{phi_resolved}
\end{equation}
Here, $S_{ij} = \frac{1}{2}  (\partial_j u_i^{\rm A} + \partial_i u_j^{\rm A})$ is the strain-rate tensor
and $W_i = u_i^{\rm B} \phi + \delta u_i\int {\rm d}{\bf x}
G({\bf x},{\bf x'}) [\partial^2_{jk} (u_j^{\rm A}\delta u_k +\delta u_j u_k^{\rm A} + \delta u_j\delta u_k)]'$.

A spatial integration with $\Phi = \int d{\bf x} \phi$ now leads to the cancellation of the divergence term and hence 
\begin{equation}
\frac{d\Phi}{dt} = - \langle \delta u_i S_{ij} \delta u_j\rangle +\nu \langle \delta u_i \nabla^2 \delta u_i \rangle + \langle \delta u_i \delta f_i\rangle \equiv \beta_S + \beta_\eta + \langle \delta u_i \delta f_i\rangle
\end{equation}
the evolution equation for our spatially integrated decorrelator as discussed in the main text of the manuscript.

% ======================================================================== %
% Beta_S
% ======================================================================== %

\subsection{Short and long-time asymptotics of $\beta_S$}
At early times, $\langle \delta u_i \delta f_i\rangle = 0$ (since the forcing terms are nearly identical), the pre saturation exponential growth of the decorrelator with the Lyapunov exponent $\lambda = \lambda^{\rm DNS} \sim \lambda_S$ 
is controlled by the short time asymptotics of $\beta_S$. Hence, in what follows, we drop the contribution from 
$\beta_\eta$, to simplify notation.   

We now expand  $\delta u_i S_{ij} \delta u_j$ in the eigenbasis of the strain-rate tensor, with eigenvalues 
$\gamma_i$ and direction cosines $\cos\theta_i$, to obtain 
\begin{equation}
	\frac{\partial}{\partial t} \phi = - \phi({\bf x},t)\, \Gamma({\bf x},t)
	\label{eq:beta_S}
\end{equation}
with 
$\Gamma({\bf x},t) \equiv \left(\sum_{i=1}^{3} \gamma_i({\bf x}, t) \cos^2\theta_i({\bf x}, t)\right) 
\approx \gamma_3({\bf x}, t) \cos^2\theta_3({\bf x}, t)$ dominated by the compressional eigendirection 
as discussed in the main text.

We choose an integrating factor  $e^{\Theta({\bf x}, t)}$, such that  
$\partial_t \Theta({\bf x}, t) = \Gamma({\bf x}, t)$, to obtain the form of the spatially-resolved decorrator 
at time $t$
\begin{equation}
\phi({\bf x}, t) = \phi({\bf x}, 0) e^{\Theta({\bf x}, 0) -\Theta({\bf x}, t)}  
	\label{eq:IF}
\end{equation}

We are now able to spatially integrate Eq.~\eqref{eq:IF} and, keeping in mind that the initially localised perturbation 
$\phi({\bf x}, 0)$ is independent of $e^{\Theta({\bf x}, 0) -\Theta({\bf x}, t)}$, we obtain
\begin{equation}
\Phi(t) = \Phi(0) \langle e^{\Theta({\bf x}, 0) -\Theta({\bf x}, t)} \rangle
	\label{eq:IF-int}
\end{equation}
where the angular brackets $\langle \cdot \cdot \rangle \equiv \int d{\bf x} (\cdot \cdot)$.

Since we are interested in the short-time asymptotics of $\beta_S$, it is possible to Taylor expand 
the exponential and, keeping in mind the mapping of $\Theta$ to $\Gamma$, we obtain
\begin{equation}
\Phi(t) \lesssim \Phi(0) \exp \left[ \sum_{p = 1}^\infty \frac{(-1)^p}{p!} \langle \Delta \Theta({\bf x}, t)^p\rangle \right]
	\label{eq:exp}
\end{equation}
where $\Delta \Theta({\bf x}, t) = \Theta({\bf x}, t) - \Theta({\bf x}, 0)$. 
Considering terms up to $\mathcal{O}(\Delta \Theta^2)$, to account for corrections stemming from intermittent 
fluctuations~\cite{BereraPRL}, 
\begin{equation}
\Phi(t) \lesssim \Phi(0) \exp \left [- \langle \Delta \Theta \rangle + \frac{1}{2}\langle \Delta \Theta^2\rangle \right ]
	\label{eq:2ndorder}
\end{equation}

Now all that remains is the evaluation of $\langle \Delta \Theta \rangle$ and $\langle \Delta \Theta^2\rangle$. 

From the choice of $\Theta$ it is easy to see (by interchanging time and space integrals 
$\langle \cdot \cdot \rangle \equiv \int d{\bf x} (\cdot \cdot)$ suitably) that $\langle \Delta \Theta \rangle \propto \langle \gamma_3 \rangle \sim  \sqrt{{\rm Re}}$ (see Fig. 4 of the main text). The determination of $\langle \Delta \Theta^2\rangle$ 
requires a further Ansatz of an approximate delta-correlation in time, at least over time-scales of the order 
associated with strain-rate, 
to yield  $\langle \Delta \Theta^2\rangle \propto \gamma_3^{\rm std} \sim {\rm Re}^{0.59}$ (see Fig. 4 
of the main text as well as Fig.~\ref{3fig}(c) which shows that indeed the distribution of 
$\gamma_3$ changes with Re) which, in the large Reynolds number limit dominates over the  $\sqrt{{\rm Re}}$ of the mean exponent. 
Hence $\lambda = \lambda^{\rm DNS} \sim \lambda_S \sim {\rm Re}^\alpha$ with $\alpha = 0.59 \pm 0.04$.

This then completes the proof of how the intermittency induced fluctuations determines the Lyapunov exponent 
for fully-developed turbulence.

At long times, beyond the exponential-growth phase when the systems have
started to completely decorrelate the forcing term can not be neglected.  Thus
as $t \to \infty$, we estimate   $\langle \delta u_i \delta f_i \rangle =
\langle u^B_i f^B_i \rangle + \langle u^A_i f^A_i \rangle - \langle (u^B_i
f^A_i + u^A_i f^B_i) \rangle \to 2\overline{\langle \varepsilon\rangle}$. As we
shall see below, this exactly compensates the contribution of the viscous term
$\beta_\eta$ at long times.

% ======================================================================== %
% Beta_eta
% ======================================================================== %
\subsection{Short and long-time asymptotics of $\beta_\eta$}

The viscous term can be expanded as 
\begin{eqnarray}
\beta_\eta &=& \nu\langle \delta u_i \nabla^2 \delta u_i \rangle = - \nu\langle \partial_j \delta u_i \partial_j \delta u_i\rangle \nonumber\\
&=& -\nu \langle(\partial_i u_j^B \partial_i u^B_j + \partial_i u_j^A \partial_i u^A_j - 2\partial_i u_j^B \partial_i u^A_j) \rangle \nonumber\\
&=& - 4\nu \operatorname{Tr}(S^2) + 2 \nu \langle \partial_i u_j^{\rm A} \partial_i u_j^{\rm B}\rangle = 2\langle \varepsilon\rangle + 2 \nu \langle \partial_i u_j^{\rm A} \partial_i u_j^{\rm B}\rangle.
\end{eqnarray}

\begin{figure}
 \includegraphics[width = 0.45\linewidth]{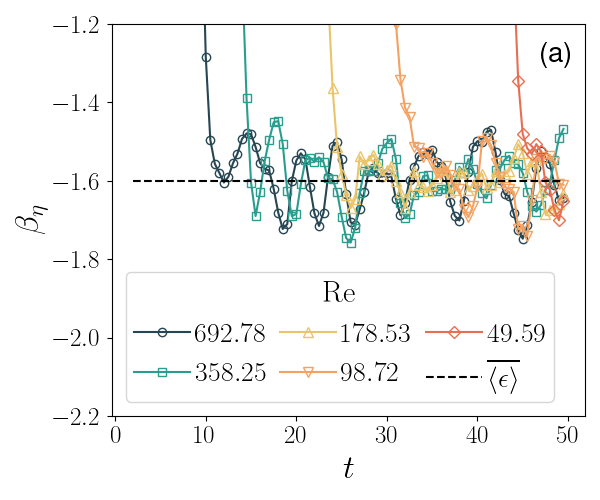}
 \includegraphics[width = 0.45\textwidth]{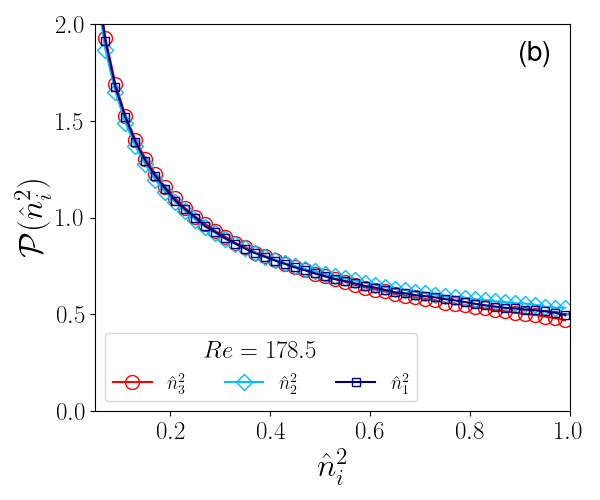}\\
 \includegraphics[width = 0.45\linewidth]{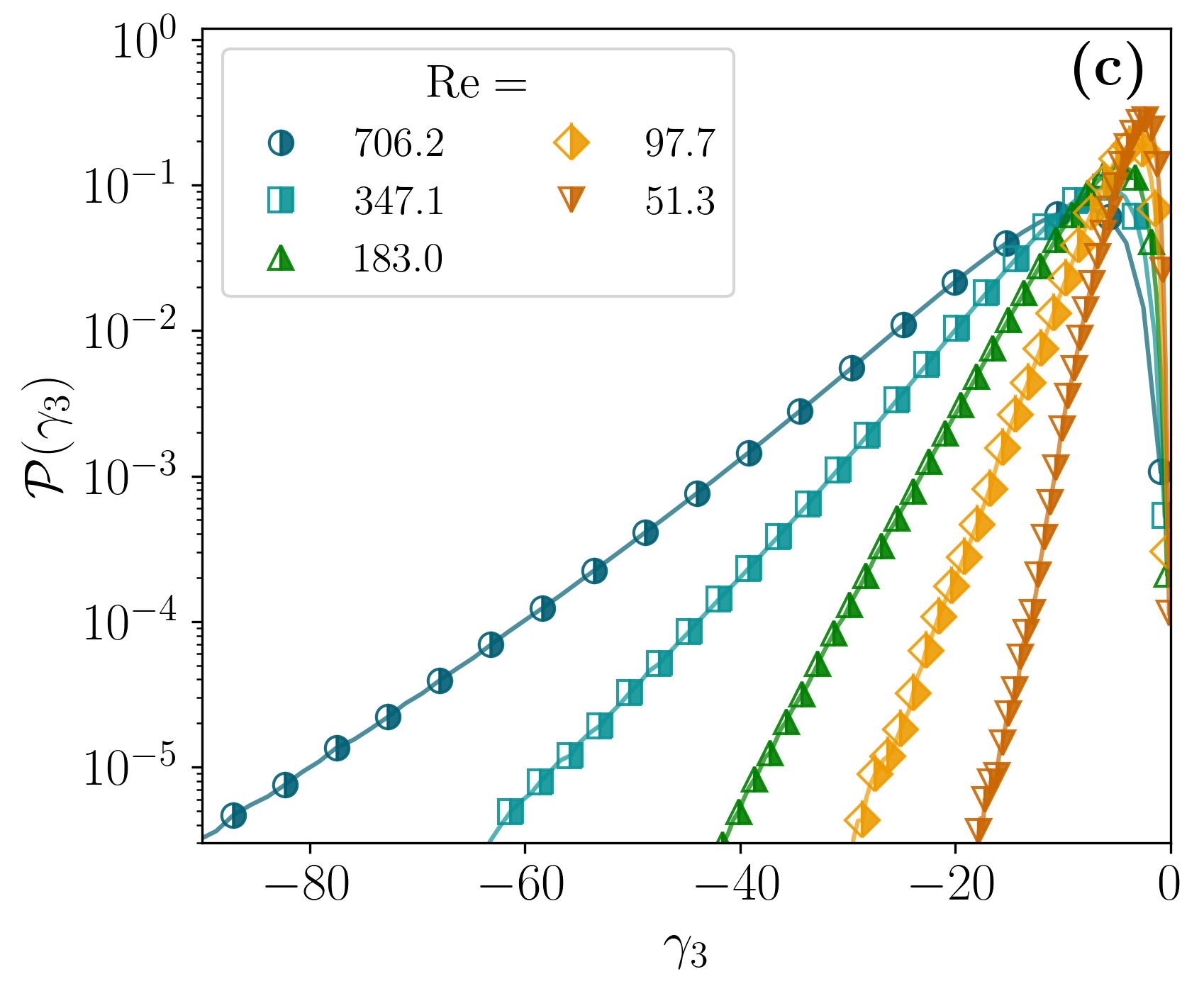}
 \includegraphics[width = 0.44\linewidth]{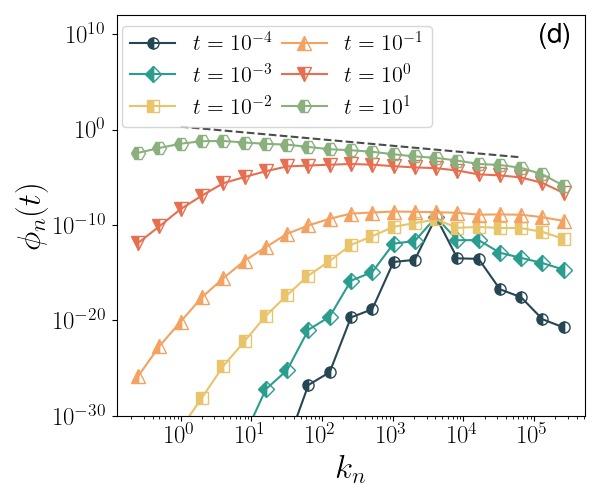}
	\caption{(a) Plots of $\beta_\eta$, at late times, 
	for different Reynolds numbers, showing a saturation to $-2 \overline{\langle \varepsilon \rangle}$ (black dotted, horizontal 
	line). (b) A plot of the probability distribution of components of $\delta {\bf u}$ (for ${\rm Re} = 358.2$), for the Navier-Stokes equation,
	along the three eigen vectors of 
	the strain rate tensor at late times once the decorrelator saturates.
	(c) The distribution of the compressional eigenvalues for different values of ${\rm Re}$. Both the peak and the spread of the distribution scales with ${\rm Re}$.
	(d) The evolution of $\phi_n$ vs $k_n$, for the GOY shell model, on a loglog scale at different times; the late 
	time shows an asymptotic scaling corresponding to $k_n^{-2/3}$ indicated by the dashed line.}
 \label{3fig}
\end{figure}

We find at short times, in the exponential-growth phase,
this terms scales as $\beta_\eta =  \mathcal{O}(1/\tau_\eta)$,
where $\tau_\eta = \sqrt{\frac{\nu}{\overline{\langle
\varepsilon\rangle} }}$ corresponds to the smallest Kolmogorov time scales. 
However, at long times $t \to \infty$, systems A and B decorrelate leading to 
$\langle \partial_i u_j^{\rm A} \partial_i u_j^{\rm B}\rangle \to 0$ and hence 
$\beta_\eta \sim -2\langle\varepsilon\rangle$ (see Fig.~\ref{3fig}(a) as well as the main text). 
This is easily understandable because, on average, dissipation is balanced by the energy injection derived above.

% ================================================================= %
% Shell model
% ================================================================= %

\section{Decorrelator in the shell model for turbulence} \label{App_shell}
\renewcommand{\theequation}{C-\arabic{equation}}
\setcounter{equation}{0}  % reset counter 
It is possible to construct an evolution equation for the decorrelator in our shell model but, as we shall see, its structure is 
less transparent than what we have seen for the Navier-Stokes equation.

By subtracting the GOY model equations for systems A and B we obtain 
the evolution of the shell-by-shell  velocity difference $\delta u_n = u_n^{\rm B} - u_n^{\rm A}$:
\begin{eqnarray}
    \frac{d}{dt} \delta u_n
    &=& \iota k_n [(\delta u_{n+2}u^{\rm A}_{n+1} + \delta u_{n+2}u^{\rm A}_{n+1} + \delta u_{n+2}\delta u_{n+1}) - \frac{\epsilon}{\lambda}(\delta u_{n+1}u^{\rm A}_{n-1} + \delta u_{n+1}u^{\rm A}_{n-1} + \delta u_{n+1}\delta u_{n-1})  \nonumber \\
    &+& \frac{(\epsilon - 1)}{\lambda^2} (\delta u_{n-2}u^{\rm A}_{n-1} + \delta u_{n-2}u^{\rm A}_{n-1} + \delta u_{n-2}\delta u_{n-1})]^* - \nu k_n^2 \delta u_n.
\label{shell_diff}
\end{eqnarray}
From this by suitably multiplying with the complex conjugate field $\delta u^*$, the decorrlator obeys 

\begin{eqnarray}
\frac{d}{dt} \phi_n &=& \frac{\iota k_n}{2} [(\delta u_{n+2}u^{\rm A}_{n+1}\delta u_{n} + \delta u_{n+2}u^{\rm A}_{n+1}\delta u_{n} + \delta u_{n+2}\delta u_{n+1}\delta u_{n}) - \frac{\epsilon}{\lambda}(\delta u_{n+1}\delta u_{n}u^{\rm A}_{n-1} + \delta u_{n+1}\delta u_{n}u^{\rm A}_{n-1} \nonumber\\
&+& \delta u_{n+1}\delta u_{n}\delta u_{n-1}) + \frac{(\epsilon - 1)}{\lambda^2} (\delta u_{n-2}u^{\rm A}_{n-1}\delta u_{n} + \delta u_{n-2}u^{\rm A}_{n-1}\delta u_{n} + \delta u_{n-2}\delta u_{n-1}\delta u_{n})]^* + c.c. - 2\nu k_n^2 \phi_n.
\label{shell_phi}
\end{eqnarray}
A summation over all shells leads to simplifications in the evolution of the GOY model decorrelator $\Phi^{\rm GOY}$ 
\begin{eqnarray}
	\frac{d}{dt}\Phi^{\rm GOY} &=& \frac{\iota}{2} \sum k_n [\epsilon\delta u_{n} u^{\rm A}_{n+1}\delta u _{n+2} + (1-\epsilon)\delta u_{n} \delta u_{n+1}u^{\rm A} _{n+2} - u^{\rm A}_{n} \delta u_{n+1}\delta u _{n+2}]^* + c.c - 2\nu \sum_n k_n^2 \phi_n
\end{eqnarray}

Further progress is only possible through a numerical solution of the
decorrelator equation as we report in the main text of the manuscript. The
shell-wise decorrelator at very short times show a power-law growth with a
wavenumber-dependent scaling exponent.  In particular at long times when the
decorrelator saturates, we find, unsurprisingly, $|\delta u_n|^2 \sim |u_n|^2
\sim k_n^{-2/3}$ as shown in Fig.~\ref{3fig}(d).

\end{document}